\documentclass[%
 reprint,
superscriptaddress,
 amsmath,amssymb,
 aps,
]{revtex4-2}

\usepackage{comment}
\usepackage{graphicx}% Include figure files
\usepackage{dcolumn}% Align table columns on decimal point
\usepackage{bm}% bold math
\usepackage{filecontents}
\begin{document}

\preprint{APS/123-QED}

\title{Strongly enhanced dynamics of a charged Rouse
dimer \\ by an external magnetic field}% Force line breaks with \\

\author{Rushikesh Shinde}
\affiliation{%
Leibniz-Institut f\"ur Polymerforschung Dresden, Institut Theorie der Polymere, K-Geb\"aude, 
Kaitzer Straße 4, 01069 Dresden, Deutschland.
}%
\affiliation{%
Technische Universit\"at Dresden, Fakult\"at Informatik, 01187 Dresden, Deutschland.}%

\author{Jens Uwe Sommer}%
\affiliation{%
Leibniz-Institut f\"ur Polymerforschung Dresden, Institut Theorie der Polymere, K-Geb\"aude, 
Kaitzer Straße 4, 01069 Dresden, Deutschland.
}%
\affiliation{
 Technische Universit\"at Dresden, Institut f\"ur Theoretische Physik, 01069 Dresden, Deutschland.
}%

\author{Hartmut L\"owen}
\affiliation{Institut f\"ur Theoretische Physik II : Weiche Materie, Heinrich-Heine-Universit\"at D\"usseldorf, 40225 D\"usseldorf, Deutschland.
}%

\author{Abhinav Sharma}
 \email{sharma@ipfdd.de}
\affiliation{%
Leibniz-Institut f\"ur Polymerforschung Dresden, Institut Theorie der Polymere, K-Geb\"aude, 
Kaitzer Straße 4, 01069 Dresden, Deutschland.
}
\affiliation{
 Technische Universit\"at Dresden, Institut f\"ur Theoretische Physik, 01069 Dresden, Deutschland.
}%

%ollaboration{CLEO Collaboration}%\noaffiliation


\begin{abstract}
While the dynamics of dimers and polymer chains in a viscous solvent is well understood within the celebrated Rouse model, the effect of an external magnetic field on the dynamics of a charged chain is much less understood. Here we generalize the Rouse model for a charged dimer to include the effect of an external magnetic field. Our analytically solvable model allows a fundamental insight into the magneto-generated dynamics of the dimer in the overdamped limit as induced by the Lorentz-force. Surprisingly, for a dimer of oppositely charged particles, we find an enormous enhancement of the dynamics of the dimer center which exhibits even a transient superballistic behavior. This is highly unusual in an overdamped system for there is neither inertia nor any internal or external driving. We attribute this to a significant translation and rotation coupling due to the Lorentz force. We also find that magnetic field reduces the mobility of a dimer along its orientation and its effective rotational diffusion coefficient. In principle, our predictions can be tested by experiments with colloidal particles and complex plasmas.

%\begin{description}
%\end{description}
\end{abstract}

\maketitle

%\tableofcontents
\section{Introduction }

The celebrated Rouse model describes the conformational dynamics of polymer chains in a viscous solvent~\cite{rouse1953theory}. The motion of a chain is usually described in terms of independently evolving Rouse modes. The slowest mode corresponds to the motion of the center of the chain which undergoes normal diffusion at all times with a coefficient that scales inversely with the number of monomers. The dynamics of the individual monomers, however, show a crossover between subdiffusive behavior at short times to normal diffusion at long times~\cite{rouse1953theory,doi1988theory}. The simplest connected structure that retains the essential physics of a polymer chain is the Rouse dimer composed of two monomers connected via a harmonic spring. While the monomers diffuse freely at short times, the long time behavior is governed by the slow moving center of the dimer. 

It seems apparent that connecting particles to form dimers or chains should result in slowing down of the dynamics. Here, we show the contrary in a system of charged particles subjected to Lorentz forces: in a dimer of oppositely charged particles, we find an enormous enhancement of the dynamics of the dimer center which exhibits even a transient superballistic behavior. Our exactly solvable model is a simple generalisation of the Rouse dimer to include the effect of an external magnetic field. We show that the strongly enhanced dynamics have their origin in the coupling of the center of mass and bond vector of the dimer due to the Lorentz force. It also follows that in longer chains, with an arbitrary charge distribution, Lorentz forces will result in correlated time evolution of the different Rouse modes. We further show that magnetic field reduces the mobility of a dimer along its orientation and its effective rotational diffusion coefficient.

The emergence of ballistic and superballistic motion in a passive, overdamped system is highly unusual for there is neither inertia nor any internal or external driving. Contrast this with driven systems where there can be strongly enhanced dynamics due to a continuous input of energy, for instance, active Brownian particles with orientation dependent mobility~\cite{ten2011brownian}, under shear flow~\cite{sprenger2020active}, diffusing in an motility landscape~\cite{breoni2020active}. Explicitly breaking detailed balance can also give rise to strongly enhanced dynamics as was shown recently in a study on random walks with a biased detachment dynamics~\cite{korosec2020apparent}. 

The unusual dynamics of the dimer can be understood in simple physical terms of Brownian diffusion under magnetic field. It is known that the curving effect of the Lorentz force leads to a reduction of the diffusion coefficient in the plane perpendicular to the applied magnetic field whereas diffusion along the direction of the magnetic field remains unaffected~\cite{lemons1999brownian, jimenez2006brownian, jimenez2008brownian}. At short times, a dimer of oppositely charged particles diffuses with a reduced diffusion coefficient as each particle diffuses independently. At long times, however, the dimer diffuses as a neutral dimer. The crossover between the two regimes--from slow magnetic field governed diffusion to magnetic field independent diffusion--is characterised by temporal exponents that can exceed 2 (superballistic) with increasing magnetic field. The long time diffusion of the dimer as a neutral particle is a telltale sign of the correlated motion of the two particles, the origin of which lies in the peculiar nature of diffusion under Lorentz force.

\begin{figure}[t]
    \includegraphics[width=\columnwidth]{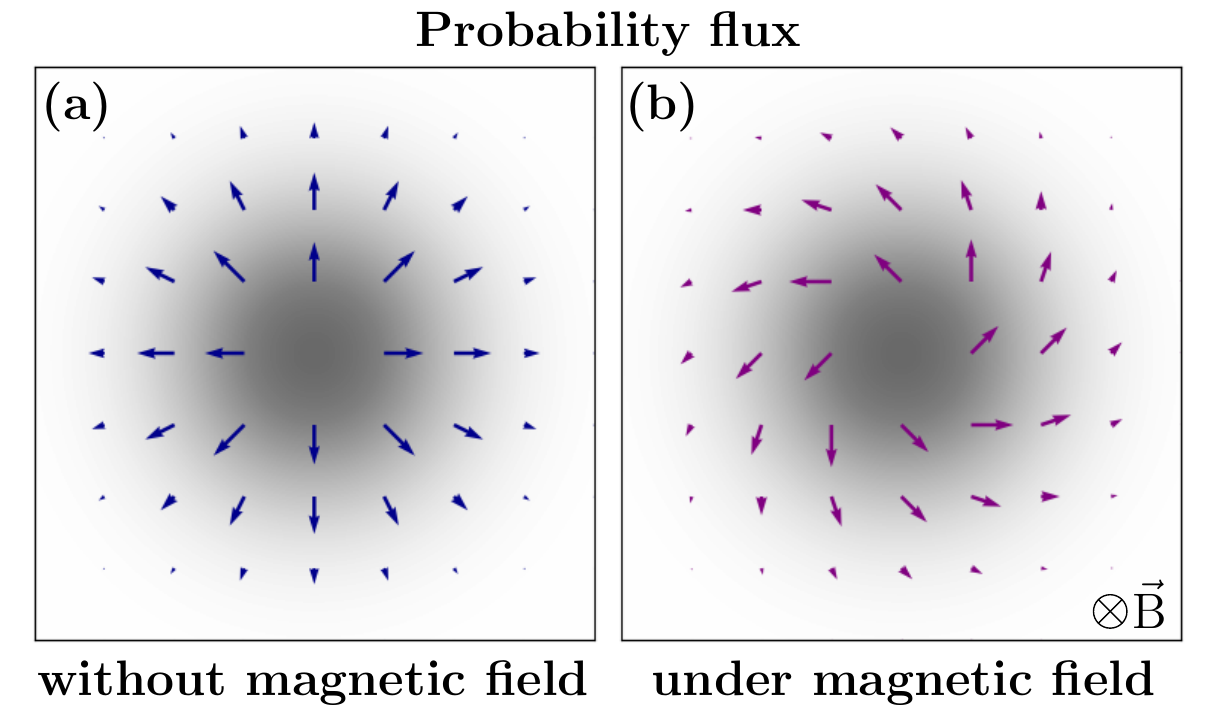}
    \caption{Probability distribution and fluxes of a particle diffusing in 2D. The particle starts the origin at $t=0$. The shaded region in (a) and (b) represents the Gaussian probability distribution at a later time. Arrows show the probability fluxes without magnetic field in (a) and with in (b). Whereas fluxes are typical of a diffusive system in (a), they are curved by the applied magnetic field in (b) which is a hallmark of Lorentz force. The curved flux can be decomposed into a diffusive flux and a rotational Lorentz flux. Note that there is a rescaling of time by a factor of $1+\kappa^2$ in (b) to account for the slower dynamics under a magnetic field.}
    \label{Lorentz_flux}
\end{figure}

\begin{figure*}
%\centering
\includegraphics[width=\textwidth]{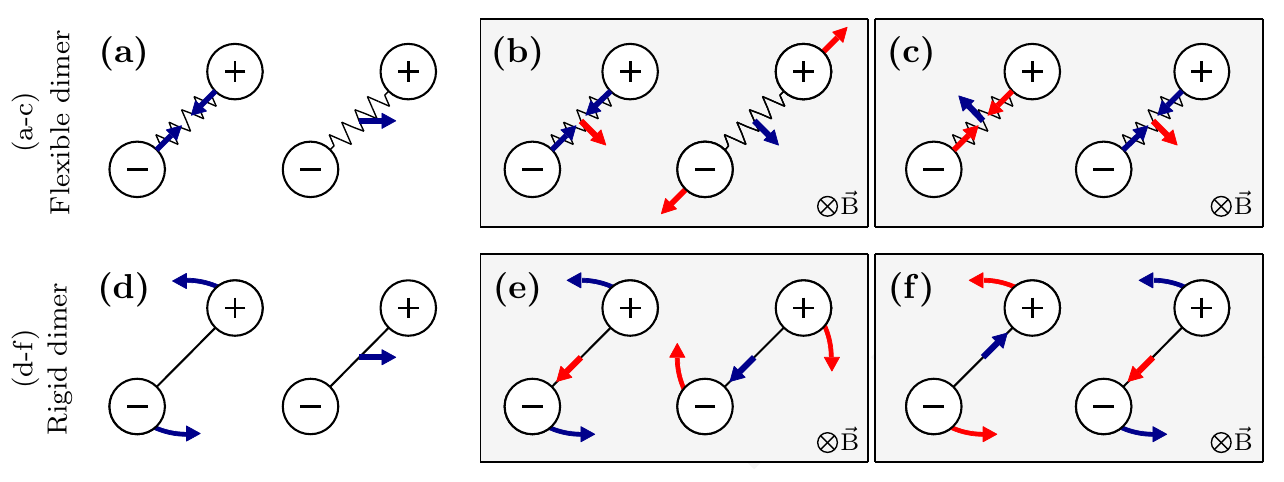}
\caption{
Dimers composed of oppositely charged particles. Blue arrows denote an imposed perturbation and the red arrow is its resultant effect. (a) In the absence of magnetic field, the motion of the bond vector and of the center of mass are uncorrelated. (a) The imposed perturbation along the bond vector does not influence the motion of the center of mass, and vice versa. Similarly, in a rigid dimer (d) the rotation and translation evolve independently of each other. However, in the presence of the Lorentz force, the motion of the bond vector gets coupled to the motion of the center of mass as shown in (b,c). (b) An imposed perturbation (in blue) along the bond vector gives rise to translation (sideways motion in red) of the center of mass. Instead, if the center of mass is translated, the resultant effect is an extension of the bond vector. (c) Translating the center of mass in the opposite direction results in contraction of the bond vector. Contracting the bond vector results in translation of the center of mass. In a rigid dimer, the Lorentz force couples the rotation of the bond vector to the translation of the center of mass as shown in (e,f). (e) Rotating a dimer results in translation of dimer along its orientation. Translating a rigid dimer along its orientation results in rotation. On reversing the direction of the imposed translation as shown in the left of (f), the resultant rotation also reverses.}
\label{schematic}
\end{figure*}

\section{Diffusion under Lorentz force}
\textbf{Diffusion Tensor.} 
The Lorentz force can significantly affect the motion of a Brownian particle when it becomes comparable to the drag force due to the viscous solvent. This is quantified by the dimensionless parameter $\kappa = qB/\gamma$, referred to as the diffusive Hall-effect parameter determined by the charge $q$ of the particle, the strength of magnetic field $B$ and the friction coefficient $\gamma$. Since the Lorentz force is perpendicular to the velocity of the particle, its motion along the direction of the applied magnetic field is not affected by the Lorentz force. In the plane perpendicular to the magnetic field, the curving effect of the Lorentz force leads to a reduction in the diffusion coefficient which is given as $D_0/(1+\kappa^2)$, where $D_0$ is the diffusion coefficient of the particle in the absence of magnetic field. 

In the absence of magnetic field, probability fluxes associated with the diffusing particle are along the density gradients according to the Fick's law~\cite{doi1988theory}. However, in the presence of a magnetic field, Lorentz force curves these probability fluxes. In Fig.~\ref{Lorentz_flux} we show the probability density and fluxes at a finite time for a particle that was initially at the origin. The motion is restricted to the $x$-$y$ plane with magnetic field along the $\hat{z}$ direction. Without magnetic field, there are only the usual diffusive fluxes (Fig.~\ref{Lorentz_flux}(a)) which can be described by the scalar diffusion coefficient $D_0 = k_BT/\gamma$, where $k_B$ is the Boltzmann constant and $T$ is the temperature. Under magnetic field, the fluxes appear curved--there are additional Lorentz fluxes perpendicular to the diffusive fluxes (Fig.~\ref{Lorentz_flux}(b)). Such diffusion cannot be captured by a single scalar coefficient. The curved fluxes are described by a tensorial coefficient $\mathbf{D} = \mathbf{D}_s + \mathbf{D}_a $ where $\mathbf{D}_s$ and $\mathbf{D}_s$ are the symmetric and antisymmetric parts of the diffusion tensor, respectively.
The diffusive fluxes are described by the symmetric tensor $\mathbf{D}_s/D_0 = \mathbf{I}/(1+\kappa^2)$ and the Lorentz fluxes by the antisymmetric tensor $\mathbf{D}_a/D_0 = -\kappa\mathbf{N}/(1+\kappa^2)$, where $\mathbf{I}$ is the identity matrix and $\mathbf{N}$ is the rotation matrix that describes $\pi/2$ rotation in the plane perpendicular to the magnetic field. Note that in an equilibrium system, the motion of a Brownian particle remains isotropic under the effect of the Lorentz force. The variance of the particle position is determined only by the even part of the diffusion tensor. The odd part of the tensor gives rise to fluxes perpendicular to the density gradients, however, it does not affect the variance of the particle position.

\noindent \textbf{Langevin equation.} An external magnetic field breaks the time-reversal symmetry~\cite{harman1963theory,balakrishnan2008elements} which is apparent in the antisymmetric components of the diffusion tensor. The broken time-reversal symmetry manifests itself in the Langevin approach via the non-white nature of noise in the equation of motion which reads as
\begin{equation}
\boldsymbol{\Gamma}  \frac{d{\vec{r}}}{dt}  = -\nabla\phi + \vec{\eta}(t),  \label{Langevin}
\end{equation}
where $\boldsymbol{\Gamma}$ is an inverse mobility tensor equal to $\gamma (\mathbf{I} + \kappa\boldsymbol{N})$ and $\phi$ is the potential. The noise $\vec{\eta}(t)$ in ~\eqref{Langevin} is a non-white Gaussian noise with zero mean and time correlation
\begin{equation}
\big \langle \vec{\eta}(t) \vec{\eta}^T(s) \big \rangle =  k_BT \left(\boldsymbol{\Gamma}^T \, \delta_+(t-s) + \boldsymbol{\Gamma} \, \delta_-(t-s)\right), \label{var}
\end{equation}
where $\delta_{\pm}(u)$ are modified Dirac $\delta$ functions. Here $\delta_+(t)$ is short hand notation for $\delta(t-\epsilon)$. $\epsilon$ is an arbitrarily small positive number. Similarly one can define $\delta_-(t)$ as $\delta(t+\epsilon)$. It immediately follows that $\delta_+(t)=\delta_-(-t)$ with $\int_0^\infty du\delta_+(u)=\int_{-\infty}^0 du \delta_-(u)=1$ and $\int_{-\infty}^0 du\delta_+(u)=\int_0^{\infty} \delta_-(u)=0$~\cite{chun2018emergence}. It was shown by Chun \emph{et. al}~\cite{chun2018emergence} that \eqref{Langevin} gives rise to the unusual fluxes as shown in Fig.~\ref{Lorentz_flux}(b).

\section{\label{Flex_dimer}Rouse Dimers}
To observe the effects of Lorentz force in an overdamped system one needs highly charged particles in a fluid medium. Examples of such systems are manifold ranging from charged colloidal suspensions \cite{hansen2000effective,masschaele2010finite} and highly charged dust particles in a complex plasma \cite{ivlev2012complex}
to vibrated granular spheres which are charged by triboelectric effects and damped from their collision with the substrate \cite{kaufman2009phase_soft,kaufman2009phase}. In order to achieve a high coupling to an external magnetic field, both high particle charges and high magnetic field strengths are needed. Actually the diffusive Hall parameter can be pushed into the order of one for granular spheres with magnetic fields of several Tesla in a low viscosity medium which is high but in reach for colloidal samples. 
Note that the magnetic field affects the motion via Lorentz forces which do not perform any work on the system. This is in contrast with the studies of ferrofluids and magnetorheological fluids where an external magnetic field interacts with the magnetic dipole moment of the particles~\cite{alvarez2013translational}.

A highly charged macroion, when put into a liquid, will attract counterions in the solution. How do the  counterions affect the diffusive behavior of the macroion? Electric bilayers formed by quasi free counterions \cite{borisov1994diagram,he2010polyelectrolyte} in a salt solution seem to have negligible effect on diffusion of a spherical particle~\cite{ohshima1984sedimentation,schumacher1987brownian}. Whether these counterions are localized close to the macroions or not, is strongly dependent on the spatial dimensionality. In three
dimensions, entropy dictates that counterions are strongly delocalized
and therefore do not contribute to screening. This is quantified
via the Debye-H\"uckel screening length, which diverges
in three dimensions in the absence of salt at high macroion dilution. This would suggest that counterion motion
in the magnetic field is probably irrelevant for our problem and
three-dimensional set-up.

To gain analytical insights, we restrict the motion of the dimer to a plane perpendicular to the direction of the magnetic field.
Dynamics in 3D and more general cases are considered in the supplementary information (SI).

\noindent \textbf{Flexible Dimers.} In the following, we do not consider an explicit Coulomb interaction between the monomers but incorporate that together with the elastic forces into an effective harmonic spring of zero rest length such that $\phi(\vec{r_1}, \vec{r_2}) = k |\vec{r}_1-\vec{r}_2|^2/2 $, where $\vec{r}_1$ and $\vec{r}_2$ are the position vectors of the two particles and $k$ is the spring stiffness. Though such a case is not viable in experiments as charges will just attract each other, the resulting model becomes analytically tractable and the results can be qualitatively extrapolated to finite rest length case. The equations of motion for the two particles are given as
\begin{align}
%\dund{\Gamma}_i
\boldsymbol{\Gamma}_1 \frac{d{\vec{r_1}}}{dt} = -k (\vec{r}_1-\vec{r}_2) + \vec{\eta}_1(t), \\
\boldsymbol{\Gamma}_2 \frac{d{\vec{r_2}}}{dt} = -k (\vec{r}_2-\vec{r}_1) + \vec{\eta}_2(t),
\label{langevin_zer0}
 \end{align}
where $i = 1,2$ is the label for the two particles and $\boldsymbol{\Gamma}_i = \gamma (\mathbf{I} + \kappa_i\boldsymbol{N})$. The noise $\vec{\eta}_i(t)$ has the time correlation as given in \eqref{var}. In what follows, we focus on the dynamics of the two fundamental Rouse modes associated with a dimer, namely, the center of mass coordinate $\vec{R} = (\vec{r}_1+\vec{r}_2)/2$ and the bond vector $\vec{r} = (\vec{r}_1-\vec{r}_2)$.

\begin{figure*}
\centering
\includegraphics[width=16cm,height=8cm]{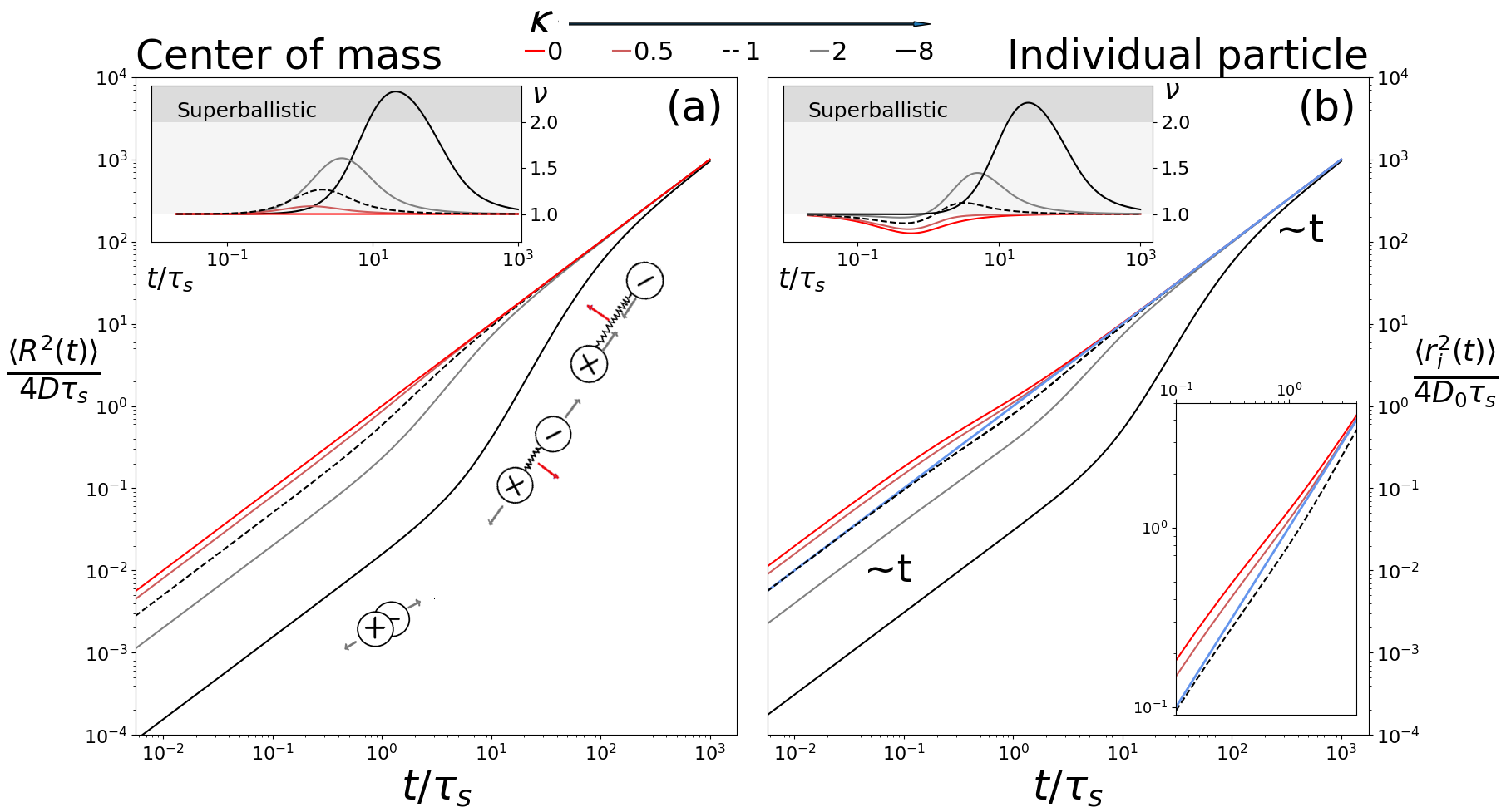}
\caption{(a) Normalised mean squared displacement (MSD) of the center of mass of a flexible dimer carrying opposite charges. The center of a dimer shows a crossover between two normal diffusive regimes: initially it diffuses slowly with a diffusion constant ${D}/(1+\kappa^2)$ because each particle diffuses independently under the influence of the magnetic field, thus extending the spring at the same time. 
As the particles move away from each other, extending the bond vector, the dimer diffuses in a direction perpendicular to its orientation. Similarly when the bond vector is shortened due to the particles approaching each other, it gives rise to Lorentz force which moves the dimer sidewise. As the bond fluctuations relax, acceleration slows down and the diffusion constant reaches the diffusion coefficient of an uncharged dimer.
(b) MSD of single particle of the dimer. Initially it has a diffusion coefficient ${D_0}/(1+\kappa^2)$ followed by a crossover into final normal diffusive regime with diffusion coefficient $D_0/2=D$. 
(Inset-a) Dynamic exponent $\nu$ is plotted as a function of $t/\tau_s$. $\nu<1$ shows subdiffusion and $\nu > 1$ indicates superdiffusion of the dimer. $\nu=2$ and $\nu>2$ are ballistic and superballistic behaviors, respectively. (Upper inset-b) Unlike center of mass, each individual particle exhibits subdiffusive behavior depending on the value of $\kappa$. (Lower inset-b) Zoomed in version of MSD for $\kappa=0,0.5$ and $1$ shows the subdiffusive behavior of individual particles. For $\kappa=1$, even though initial and final diffusion coefficients are equal, particle shows rich dynamics. Blue line shows MSD of uncharged particle with diffusion constant $D = D_0/2$.
}
\label{fig:MSD}
\end{figure*}

We first consider the case of a dimer carrying equal charges on both ends, i.e., $\boldsymbol{\Gamma}_1=\boldsymbol{\Gamma}_2=\boldsymbol{\Gamma}$. The Langevin equations for $\vec{R}$ and $\vec{r}$ are decoupled and read as
 \begin{equation}
     \frac{d\vec{R}}{dt}=\frac{1}{2} \boldsymbol{\Gamma}^{-1},  \vec{\zeta}\hspace{20pt} \frac{d\vec{r}}{dt}=\boldsymbol{\Gamma}^{-1} \Big( \vec{\xi}  - 2 k \vec{r} \Big),
     \label{eq:qq_dimer}
 \end{equation}
 where $\vec{\zeta}$ and $\vec{\xi}$ are Gaussian non-white noises with zero mean and twice the variance mentioned in \eqref{var}. Mean squared displacement (MSD) for the center of mass and length of the bond vector are
 \begin{equation}
     \left \langle R^2 \right \rangle=\frac{2D_0}{1+\kappa^2}t, \hspace{20pt} \left \langle r^2 \right \rangle = 2D_0\, \tau_s \, \Big( 1-e^{-\frac{4t}{\tau_m}} \Big).
 \end{equation}
 It follows that the diffusion of the center of mass is equivalent to the diffusion of a particle having charge $2q$ and diffusion coefficient $D_0/2$.
%where $D_i =  \frac{\gamma}{k} \, (1-e^{-\frac{4\mut}{\gamma}}) $ and
 In the above expression, the natural spring relaxation time is denoted by $\tau_s$ and is equal to $\gamma/k$. Under the influence of a magnetic field the spring relaxation time increases to $\tau_m$ which reads as
\begin{equation}
\tau_m = (1+\kappa^2)\tau_s.
\label{eq:magnetic_timescale}
\end{equation}
 \par

We now consider a dimer composed of oppositely charged particles, i.e.,  $\boldsymbol{\Gamma}_1=\gamma (\mathbf{I}+\kappa \mathbf{N} )$ and $\boldsymbol{\Gamma}_2=\gamma (\mathbf{I}-\kappa \mathbf{N} )$. The equations for $\vec{R}$ and $\vec{r}$ are given as
\begin{equation}
\frac{d\vec{R}}{dt} = \frac{1}{\gamma (1+\kappa^2)} \Big( \vec{\zeta}  + \kappa \mathbf{N} \vec{\xi} - 2k \kappa \mathbf{N} \vec{r}   \Big) \label{Collective},
 \end{equation}
\begin{equation}
\frac{d\vec{r}}{dt}  = \frac{1}{\gamma (1+\kappa^2)} \Big( -\kappa \mathbf{N} \vec{\zeta} + \vec{\xi} - 2k\vec{r}    \Big). \label{Inner}
 \end{equation}
Comparing with \eqref{eq:qq_dimer}, one can infer from \eqref{Collective} that the movement of the center of mass gets a contribution from the bond vector fluctuations. Similarly, \eqref{Inner} indicates that the dynamics of the bond vector are also dependent on the movement of the center of mass. As a result, translation of the center of mass causes a change in the length of the bond vector. Likewise, bond vector fluctuations induce fluctuations in the position of the center of the mass. The schematic shown in Fig.~\ref{schematic}(a,b) describes the coupling between the dimer center and the fluctuations of the connecting bond vector. Note that the Lorentz force induced coupling only affects the dynamics of the dimer without altering the equilibrium properties. This is in strong contrast to systems in which coupling has been induced by activity, as in active-passive dimers~\cite{vuijk2021chemotaxis} and mixtures of particles in contact with different thermostats~\cite{grosberg2015nonequilibrium}.

The above equations can be transformed to obtain a set of equations that constitute a multivariate Ornstein Uhlenbeck process. Thus one can exactly obtain the joint probability distribution of $\vec{R}$ and $\vec{r}$ as shown in the SI, section 2. 
The expression for the MSD of the center of mass reads
 \begin{equation}
\small
\frac{ \left \langle R^2(t) \right \rangle}{4 D \tau_s} = \frac{ \kappa^2}{4}\left( 1 - e^{-\frac{4t}{\tau_m}} \right) -  \kappa^2 \left( 1 - e^{-\frac{2t}{\tau_m}} \right)  + \frac{t}{\tau_s} \label{MSD},
\end{equation}
where, $D = D_0/2$ is the diffusion constant of the center of mass of an uncharged dimer. MSD for the length of the bond vector is
 \begin{equation}
\frac{\left \langle r^2(t) \right \rangle}{4D \tau_s} = 1 - e^{-\frac{4t}{\tau_m}} .
\end{equation}
As shown in Fig. \ref{fig:MSD}, the dimer center exhibits normal diffusion both at short and long times albeit with different diffusion coefficients. While the initial diffusion is governed by the magnetic field, at long times the diffusion coefficient of the dimer becomes equal to that of an uncharged dimer. To gain insight into the crossover between the two diffusive regimes, we write short time expansion of the \eqref{MSD}.
\begin{equation}
\frac{\left \langle R^2(t) \right \rangle}{4D \tau_s} =  \frac{t}{\tau_m} + \frac{4k^2}{3}\left(\frac{t}{\tau_m}\right)^3-2k^2\left(\frac{t}{\tau_m}\right)^4 + \mathcal{O}(t^5) .
\end{equation}
The above expansion shows that initially the center of mass undergoes normal diffusion with diffusion constant $D/(1+\kappa^2)$. This regime persists till time ${(1/\kappa}+\kappa)\tau_s$. The emergence of this time scale can be attributed to the spring force term in \eqref{Collective}, which contributes to the diffusion of the center of mass. On the time scale $\tau_m$, the diffusion constant of the center of mass reaches $D$. Since $D>D/(1+\kappa^2)$ and both initial and final diffusive regimes are normal, it follows that the center of mass has to undergo accelerated motion. One can quantify the acceleration by calculating the dynamic exponent $\nu=d \langle \textrm{ln}(R^2(t)) \rangle / d \textrm{ln}(t)$ as shown in the Fig. \ref{fig:MSD}. Surprisingly, the exponent can exceed $2$ for sufficiently large $\kappa$--a signature of superballistic behavior. This speed up in the crossover is unusual given that the dynamics are fully overdamped and the system is not driven out of equilibrium.

Like the center of mass, individual particles also exhibit rich dynamics.
Both particles have equal MSD, i.e., $ \langle r_1^2(t) \rangle = \langle r_2^2(t)  \rangle$, which reads as 
\begin{equation}
\small
    \frac{\left \langle r_i^2(t) \right \rangle}{2D_0 \tau_s} =  \frac{\kappa^2+1}{4} \left( 1 - e^{-\frac{4t}{\tau_m}} \right) + \frac{t}{\tau_s} - \kappa^2\left( 1 - e^{-\frac{2t}{\tau_m}} \right).
\end{equation}
The short time expansion of above expression gives 
\begin{equation}
    \small
    \frac{\left \langle r_i^2(t) \right \rangle}{4D_0 \tau_s} =    \frac{t}{\tau_m} -\left(\frac{t}{\tau_m}\right)^2 + \frac{4}{3}\left(\frac{t}{\tau_m}\right)^3+ \frac{2 \kappa^2}{3}\left(\frac{t}{\tau_m}\right)^3 + \mathcal{O} (t^4).
    \label{short_time_mag}
\end{equation}
If no magnetic field is present, the above expression reduces to
\begin{equation}
    \frac{\left \langle r_i^2(t) \right \rangle}{4D_0 \tau_s} =  \frac{t}{\tau_s} -\left(\frac{t}{\tau_s}\right)^2 + \frac{4}{3}\left(\frac{t}{\tau_s}\right)^3 + \mathcal{O} (t^4). 
    \label{uncharged_single_particle}
\end{equation}
The above expression shows that the diffusion constant of a particle is initially $D_0$. Thereafter, due to the spring forces, the particle experiences deceleration until it reaches the diffusion constant $D_0/2$. Negative terms in the above short time expansion are a telltale sign of this. Comparison of the above expression with \eqref{short_time_mag} shows that the magnetic field slows down the dynamics by the factor of $(1+\kappa^2)$ as the spring relaxation time $\tau_s$ is now changed to $\tau_m$.  Also note, apart from the rescaling of the time scale, \eqref{short_time_mag} has an additional positive term that is not present in \eqref{uncharged_single_particle}. This indicates that the Lorentz force also makes an additional contribution to the MSD of a particle.  One can easily show that these extra odd order terms arise because of the presence of anti-symmetric components of the diffusion tensor. Initially, the magnetic field reduces the diffusion constant of a particle to $D_0/(1+\kappa^2)$. For small $\kappa$, this diffusion constant will be greater than $D_0/2$. Thus, on average, particle has to experience deceleration. Therefore, in Fig. \ref{fig:MSD} for small $\kappa$, the dynamic exponent $\nu$ is below unity. As $\kappa$ increases, the Lorentz flux contribution to the MSD increases. This explains why for high values of $\kappa$, particles do not undergo decelerated motion. 

\begin{figure}[]
\centering
\includegraphics[width=8.4cm,height=7.9cm]{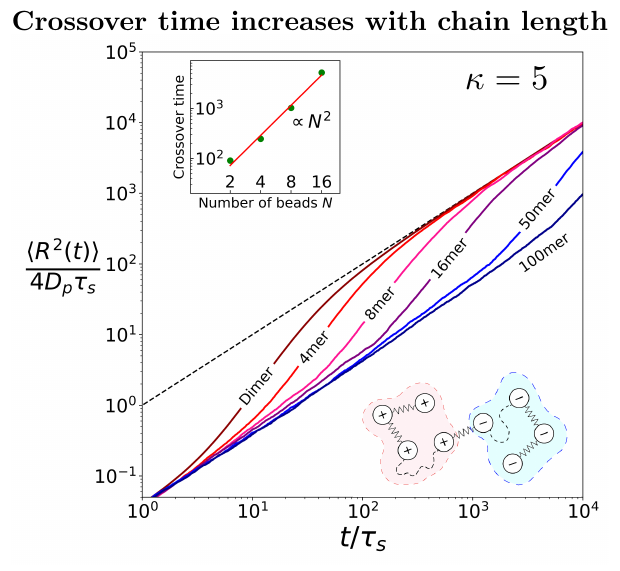}
\caption{(Inset) Schematic shows a polymer-like chain of $N$ beads. The chain is overall charge neutral with two blocks of equal and opposite charges. Charged beads are connected by springs and the broken line indicates a succession of many bead/springs. (Plot) Similar to a flexible dimer, the center of mass of a chain also shows a crossover  between two normal diffusive regimes: initially with a diffusion constant $D_{p}/(1+\kappa^2)$ and later with diffusion constant of $D_{p}$. Here $D_{p}=D_0/N$ is the diffusion constant of the center of mass of an uncharged chain. The crossover also widens and shifts forward in time with increasing chain length.}
\label{fig:MSD_polymer}
\end{figure}

While charged Rouse dimers serve as the simplest model of polyelectrolyte dynamics in a magnetic field, a better understanding of the dynamics requires longer chains with arbitrary charge distributions. In this case, Lorentz forces will induce correlated time evolution of the different Rouse modes. How does the size of Rouse chain affect the crossover period? Here we specifically consider the case of a Rouse chain with block distributed charges such that half of the chain is positively charged and the other half is negatively charged. Figure \ref{fig:MSD_polymer} shows the MSD of the center of mass of the Rouse chain. Data for other charge distributions are shown in the SI. While longer chains exhibit qualitatively the same enhancement as a dimer, the effect is much more pronounced: the onset of enhancement as well as the crossover time increase dramatically with increasing charges. Specifically, the width of the crossover region scales as $N^2$, where $N$ is the number of beads in the chain. Note that this corresponds to the slowest mode of relaxation (Rouse time) of the polymer chain.

%%%%%%%%%%%%%%%%%%%%%%%%%%%%%%%%%%%%%%%%%%%%%%%%%%%%%%%%%%%%%%%%%%%%%%%%%%%%%%%%5
%%%%%%%%%%%%%%%%%%%%%%%%%%%%%%%%%%%%%%%%%%%%%%%%%%%%%%%%%%%%%%%%%%%%%%%%%%%%%%%%5

\noindent\textbf{\label{rigid}Rigid Dimers.} We study magnetic field induced coupling of rotation and translation of a rigid dimer and its influence on the diffusive motion. The dimer consists of two particles connected via a rigid bond of length $2r_0$. The translational and rotational diffusion constants of an uncharged dimer are given as $D_t = k_BT/\gamma_t$ and $D_r = k_BT/\gamma_r$, respectively, where $\gamma_t = 2\gamma $ and $\gamma_r =2\gamma r_0^2$.
The two ends of the dimer carry charges $q_1$ and $q_2$. The unit vector along the bond joining these two charges is denoted by $\hat{u}$.

A dimer translating with a velocity $\vec{\vartheta}$ and rotating with angular velocity $\vec{\Omega}$ experiences Lorentz force
\begin{equation}
\vec{F}_{L} = -2q^+B \mathbf{N}  \vec{\vartheta} + 2 q^- B r_0 \mathbf{N} \mathbf{U}   \vec{\Omega} \label{force}
\end{equation}
and Lorentz torque
\begin{equation}
\mathcal{T}_{L}=- 2 q^- B r_0 \mathbf{U} \mathbf{N}  \vec{\vartheta},  \label{torque}
\end{equation}
where $q^+= (q_1+q_2)/{2}$ and $ q^-= (q_1-q_2)/{2}$. $\mathbf{U}$ is cross product matrix such that $\hat{u} \times (..) = \mathbf{U} \cdot (..)$. Consider a dimer of opposite charges, i.e., $q^+ = 0$. The coupling between the translational and rotational motion of the dimer is evident in the equations above (\eqref{force} and \eqref{torque}): Whereas the Lorentz force depends on the rotational velocity the Lorentz torque is determined by the translational velocity (see Fig. \ref{schematic}). Physically this implies that rotation results in Lorentz force along the orientation of the dimer. Similarly, translating the dimer along its orientation simultaneously makes it rotate.

Exactly solving the Langevin equations of rigid dimer to obtain the probability density function for spatial and orientational distribution is a formidable task. Hence we focus on deriving the corresponding Fokker-Planck equation (FPE) \cite{dhont2004rod} and deduce the diffusion behaviour from it (SI, section 3). It is convenient to define the parameters $\mathcal{K}$, $\alpha$ and $\beta$ as follows
\begin{equation}
\mathcal{K}=2q^-Br_0, \hspace{10pt} \alpha=  \frac{2q^-Br_0}{\sqrt{\gamma_r \gamma_t}}, \hspace{20pt}    \beta= \frac{2 q^+ B}{\gamma_t} .
\end{equation}
For a rigid dimer $\alpha=(q_1-q_2)B/2\gamma$ and $\beta= (q_1+q_2) B/2\gamma$. The parameter $\beta$ is analogous to the diffusive Hall-effect parameter $\kappa$. It is the ratio of the total Lorentz force to the net viscous drag on the dimer. The parameter $\alpha$ denotes the strength of coupling between rotational and translational degrees of freedom which is possible only if $q_1\neq q_2$.

We first consider the case of a rigid dimer with equal charges, i.e., $ \hspace{5pt} \alpha=0$ and $\beta=qB/\gamma$. We use superscript $+$ to indicate that the corresponding tensors are associated with a dimer carrying equal charges. Since $\alpha = 0$, there is no coupling between rotational and translational motion. The governing FPE is given as
\begin{equation}
\frac{\partial P}{\partial t} =  \nabla \cdot \Big [ \mathbf{D}^+_t \,  \nabla P \Big] + \hat{\mathcal{R}} \cdot \Big [ {D}^+_r \, \hat{\mathcal{R}} \,P \Big] \label{eq:FP_eq},
\end{equation}

where,
\begin{equation}
\mathbf{D}^+_t = D_t
\begin{bmatrix}
\vspace{5pt}
\frac{1}{1 + \beta^2}  & \frac{\beta}{1 + \beta^2} \\
\vspace{5pt}
 \frac{-\beta}{1 + \beta^2} & \frac{1 }{1 + \beta^2}
\end{bmatrix},
\hspace{20pt}
D^+_r= D_r.
\label{rigid_equal}
\end{equation}
$\hat{\mathcal{R}}$ is a rotation operator defined as $\hat{u} \times \nabla_{\hat{u}}$ which in polar coordinates is equal to $\partial/\partial \theta$.

Similar to the case of a flexible dimer, \eqref{rigid_equal} shows that the translational diffusion behaviour is identical to the motion of a particle carrying charge $2q$ and friction coefficient $2\gamma$ whereas the rotational diffusion of the dimer remains unaffected. 
Interestingly, we find that in 3 dimensions the rotational diffusion tensor becomes anisotropic and has asymmetric components (SI, section 4). This implies that the magnetic field does not only introduce Lorentz fluxes in the real space but also in orientation space.

\begin{figure}[t]
    \centering
    \includegraphics[width=0.48\textwidth]{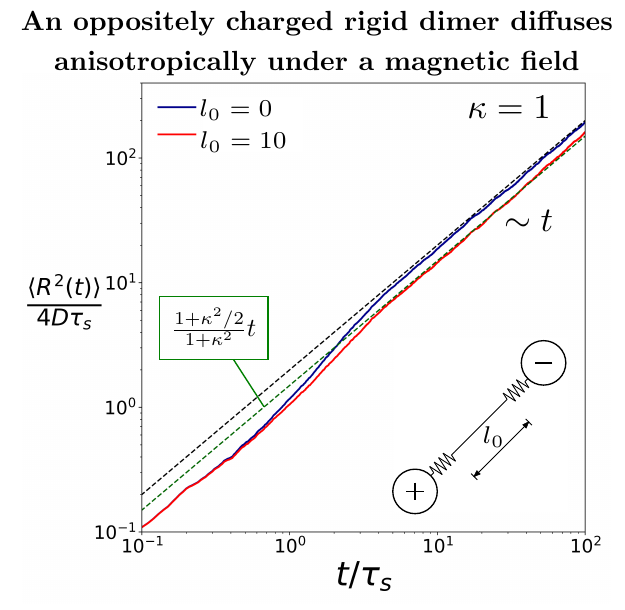}
    \caption{(Inset) Schematic of a dimer with finite rest length. The dimer is composed of two oppositely charged beads connected by a spring of $k = 1$ and rest length $l_0$. (Plot) The normalised mean squared displacement, obtained by Brownian dynamics simulations, is shown for two cases: a Rouse dimer, i.e., a dimer with a rest length of zero, and a flexible dimer with a rest length of $10 \sqrt{k_BT/k}$.  Initially, the flexible dimer with a non-zero rest length behaves like a Rouse dimer. Both dimers show enhancement in their dynamics but their final diffusion coefficients are not the same. At long times, the diffusion constant of a dimer with non-zero rest length is equal to $(1+0.5\alpha^2)D/(1+\alpha^2)$. This is the diffusion constant of a rigid dimer carrying opposite charges. These simulation results corroborate the analytical prediction that a rigid dimer carrying opposite charges diffuses anisotropically under magnetic field. These results are shown for $\alpha=1, \kappa=1$.
    }
    \label{fig:non-zero}
\end{figure}

A much more interesting case is that of a dimer carrying opposite charges, i.e., $\alpha=qB/\gamma$ and $\beta=0$. Below, a $-$ superscript is used to denote that the dimer carries opposite charges. The governing FPE is
\begin{align}
\frac{\partial P}{\partial t} =&  \nabla \cdot  \mathbf{D}^-_t  \Big [\,  \nabla P + \frac{\mathcal{K}}{\gamma_r} \hat{\mathcal{R}} P \, \, \hat{u} \Big] + \notag\\&\hat{\mathcal{R}} \cdot D^-_r \Big [  \, \hat{\mathcal{R}}P -\frac{\mathcal{K}}{\gamma_t} \hat{u}\cdot \nabla P \Big]. \label{opp_FP} 
\end{align}
Straightway one can observe that the above FPE equation has two extra terms that are not present in \eqref{eq:FP_eq}. These cross-coupling terms are the manifestation of the correlated dynamics of the dimer. These terms can give rise to extra probability fluxes which can be diffusive and hence relevant for the time evolution of probability. This is obvious if we refer to the Figure \ref{schematic}. Movements in real space that are analogous to spatial gradients of probabilities cause fluxes in orientation space. Similarly, rotational gradients will affect how probability function evolves in real space. In the above FPE,
\begin{equation}
    \mathbf{D}^-_t \big( \hat{u} \big)=D_t \left[ \mathbf{I}-\frac{\alpha^2}{1+\alpha^2}\mathbf{\hat{u}\hat{u}} \right] \label{eq:diff_opp}.
\end{equation}
On time scales greater than the relaxation of orientational degree of freedom, $\mathbf{D}^-_t\big( \hat{u} \big)$ can be regarded as the diffusion tensor for a dimer that is oriented along $\hat{u}$. This diffusion tensor is always symmetric which implies that there are no Lorentz fluxes in real space. Surprisingly, this tensor is equal to $D_t\mathbf{I}$ only when $\alpha=0$. Because of the finite separation between charges, the magnetic field reduces the diffusion constant. This is in contrast to the flexible dimer case, in which the long time diffusive behavior is independent of the magnetic field. 

A magnetic field also slows down the rotational diffusion coefficient $D^-_r$ which is given as
\begin{equation}
D^-_r=
 \frac{D_r}{1+\alpha^2}.
\end{equation}
This result seems similar to the case of a single charged particle, where dynamics in the translation space get slower by a factor of ${1}/(1+\kappa^2)$ as the magnetic field curves the trajectory of a particle. Does this mean that slower rotational diffusion is a consequence of rotation caused by a magnetic field in rotational space?  We calculate the general rotational diffusion tensor in 3 dimensions and show that it is a symmetric tensor (SI, section 4). This implies that the magnetic field does not decrease rotational mobility by precessing the orientation around some permanent axis.

The origin of the reduced coefficient is in the coupling of rotational and translational motion and can be understood as follows. Consider the scenario shown in Fig. \ref{schematic}(c). If no magnetic field is present, a dimer rotating with a constant angular velocity $\Omega$, experiences a torque of magnitude $\gamma_r\Omega$. However, due to the coupling of rotation and translation, as the dimer rotates, the magnetic field pushes the dimer along its orientation with the velocity of magnitude $\mathcal{K}\Omega/\gamma_t$. This results in a counter-torque of magnitude $\mathcal{K}^2\Omega/\gamma_t$ (See Fig \ref{schematic}). As a result, the torque $\mathcal{T}$ required to keep the dimer rotating at $\Omega$ is
\begin{equation} 
\mathcal{T} = \gamma_r(1+\alpha^2){\Omega},
\end{equation}
which is $(1+\alpha^2)$ times that of the torque needed to rotate the dimer in the absence of a magnetic field. Thus, it is evident that the magnetic field increases the rotational drag on a particle and in turn slows its rotations down.

Similarly, one can motivate the physical reasoning behind \eqref{eq:diff_opp} which indicates that the diffusion depends on the orientation of the dimer even though the diffusion tensor of an uncharged dimer is isotropic. If the dimer moves with a velocity $\vec{\vartheta}$, its component along the orientation $\hat{u}$ gives rise to a force that rotates the dimer with an angular velocity of magnitude $-\frac{\mathcal{K}}{\gamma_r}\hat{u} \cdot\vec{\vartheta}$. As discussed previously, this induced rotation propels the dimer along the orientation with the force $-\mathcal{K}^2\mathbf{\hat{u}\hat{u}}\vec{\vartheta}/\gamma_r$ (See Fig \ref{schematic}).
Hence, to keep the dimer moving  with the velocity $\vec{\vartheta}$, the required force is
\begin{equation}
    \vec{F} = \gamma_t(\mathbf{I}+\alpha^2\mathbf{\hat{u}\hat{u}})\vec{\vartheta} \label{eq:mag_drag}.
\end{equation}
The inverse of the tensor $\gamma_t[\mathbf{I}+\alpha^2\mathbf{\hat{u}\hat{u}}]$ is indeed the diffusion tensor $\mathbf{D}^-_t$. It implies that the required force to translate the dimer along its orientation is $(1+\alpha^2)$ times that of the force needed to translate it along its orientation in the absence of a magnetic field. Therefore, the diffusion along the orientation of the dimer gets reduced by the factor of ${1}/(1+\alpha^2)$ but diffusion perpendicular to $\hat{u}$ remains unaffected. Hence, the resultant diffusion tensor in \eqref{eq:diff_opp} is anisotropic. Brownian dynamics simulation results also agree with above prediction that a rigid dimer carrying opposite charges diffuses anisotropically under the influence of magnetic field  (SI, section 4). As $\alpha$ increases, the mobility of the dimer along its orientation reduces drastically but the dimer can still freely diffuse in the direction perpendicular to its orientation vector. We note that whereas the ratio of translational and rotational diffusion is a geometric constant, this is no longer the case in the presence of a magnetic field.

It is interesting to consider the effect of a high magnetic field on the behavior of rigid dimers. Note that in the limit of large magnetic fields, the oppositely charged rigid dimer can not rotate and is immobile along its orientation. It can only diffuse along the direction perpendicular to its orientation. While a charged Brownian particle is immobile under the influence of large magnetic fields as its diffusion constant is zero, on pairing two immobile particles of opposite charges via a rigid bond, one obtains a mobile dimer. This is analogous to fractons \cite{nandkishore2019fractons}. Isolated fractons are immobile quasiparticles but can be mobile by forming bound states \cite{pretko2020fracton}. The similarity between fractons and rigid dimers is evident. A single particle under high magnetic field is like a fracton and oppositely charged rigid dimer is a lineon due its restricted mobility along its orientation \cite{doshi2021vortices}.

Finally, we comment on the general case of a dimer carrying arbitrary charges. Interestingly, even for arbitrarily high charges, translational diffusion of the dimer does not go to zero. It can also be shown that unless $\alpha = 0$, the diffusion behaviour of dimer carrying charges $q_1$ and $q_2$ will not be similar to the diffusion behaviour of a single particle carrying net charge $q_1+q_2$. This suggests, due to the effect of a magnetic field,  one can not treat a dimer as a single particle carrying some equivalent charge as this can not accurately capture the diffusive dynamics of the dimer. We have also extended this analysis to rod like particles (SI, section 3, 5 and Fig S1 ).

\section{Discussion and Outlook} \label{sec4}

Onsager reciprocal relations imply that diffusion tensors are symmetric. However, the validity of these relations is limited to systems that respect time-reversal symmetry \cite{onsager1931reciprocal}. The motion of a charged particle in a magnetic field is a classical example of broken time-reversal symmetry. For this reason, the probability flux associated with a particle performing overdamped motion gets a certain handedness based on the charge it carries. Such a diffusion behaviour is described by a tensor that has anti-symmetric components. Analogous to case of odd-viscosity \cite{avron1998odd,banerjee2017odd,reichhardt2021active}, recently such diffusive behaviour has aptly been termed as odd-diffusive \cite{hargus2021odd} and has attracted considerable attention \cite{matsuyama2021anomalous,chun2019effect,park2021thermodynamic}.

Usually, under the influence of an external magnetic field, the diffusion time scale of a particle gets enhanced by a factor of $1+\kappa^2$. We showed that for a dimer, the dynamics are indeed slowed down but only at short times. There exists a time scale, determined by the magnetic field and the spring constant, beyond which the center of an oppositely charged dimer exhibits strongly enhanced dynamics which, transiently, can be even superballistic. The origin of the accelerated dynamics is the coupling between the motion of the center of mass and the bond vector fluctuations. At long times, the dimer diffuses as an ordinary neutral Rouse dimer. In principle, if the time of crossover between the two regimes--from slow magnetic field governed diffusion to magnetic field independent diffusion--is long enough, one can deduce information about the strength of bond from the mean square displacement of the center of mass. 
 On a phenomenological level, the enhanced dynamics of charged Rouse dimers under magnetic field is similar to active particles, which also exhibit crossover between two diffusive regimes \cite{marchetti2013hydrodynamics,bechinger2016active}.
However, despite this similarity, charged Rouse dimers under magnetic field cannot be regarded as an example of active systems. While active particles violate fluctuation dissipation relation, the motion of a charged Rouse dimer is governed by equilibrium fluctuations.

We now discuss an accessible experimental realisation where enhanced dynamics of charged dimers and chains can be demonstrated. Lorentz force significantly affects the motion of a colloidal particle when the diffusive Hall-effect parameter becomes comparable to unity. Let us consider $\kappa = qB/6\pi \eta R$, where $B$ is the magnetic field, $\eta$ is the viscosity of the fluid, $q$ is the total charge on the surface of a sphere of radius $R$. For a surface charge density $\sigma = 1$ nm$^{-2}$, the viscosity $\eta \approx 10^{-4}$ Pas (Propylene at room temperature), $B = 5$ T and a millimeter sized sphere where inertia still can (almost) be ignored one obtains $\kappa \approx 5$. For this $\kappa$, the dynamics are strongly enhanced as shown in Fig. \ref{fig:MSD_polymer}. Moreover, since the crossover time for a Rouse chain with $N$ beads scales as $N^2$, the observation of enhanced dynamics could be experimentally accessible for long chains. Admittedly, to obtain this value of $\kappa$, we considered large magnetic fields, low viscosity and large particles. However, these are, in principle, experimentally accessible. Millimeter-sized granules beget high charges (with both signs) when exposed to a vibrating substrate due to triboelectric effects \cite{kaufman2009phase_soft,kaufman2009phase} and therefore could provide a macroscopic realization for a model Rouse polymer. 

A plausible experimental setup can also be realised in dusty plasma. They can be almost overdamped for the high density of the ambient gas. Large magnetic fields can be effectively realised in non-inertial rotating frames; then the Coriolis force acts as the Lorentz force due to an external magnetic field. This concept has been used to magnetize complex plasma giving rise to effective magnetic fields exceeding $10^4$ T~\cite{kahlert2012magnetizing,hartmann2019self}.
In fact, millimeter-sized dust particles in a complex plasma can exhibit large electric dipole moments~\cite{hou2018structures} which can be exposed to large magnetic fields such that a Hall parameter $\kappa$ of about unity is in reach.
However, it is not yet clear how one would prepare oppositely charged particles in the rotating electrode setup.

In contrast to a flexible dimer, the long-term diffusion behaviour of a rigid dimer carrying opposite charges depends on the strength of a magnetic field. We found that the magnetic field reduces the mobility of a rigid dimer along its orientation and its effective rotational diffusion coefficient. We attribute this to the strong coupling between translation and rotation due to the Lorentz force.  This is strikingly evident in its diffusive behaviour--it does not diffuse as a neutral Rouse dimer. The reduction of rotational diffusion occurs due to the counter-torque generated by the translational motion of the dimer. In other words, if the dimer is fixed at its center and allowed to rotate freely, magnetic field will not affect its rotation. It is interesting to compare this with a recently developed method of tuning rotational diffusion in which a randomly fluctuating magnetic field was used to exert torque on active Brownian particles carrying a magnetic moment~\cite{fernandez2020feedback}.

Our work could be relevant to fractons which are particles that are not mobile on their own but they can become mobile when paired \cite{nandkishore2019fractons, schmidt2019light}. The analogy to charged Rouse dimers under magnetic field is striking: the diffusion coefficient of a single charged particle vanishes in the limit of large magnetic fields. However, when paired with another charged particle, for instance, with a particle of opposite charge via a rigid bond, the dimer becomes mobile; it can translate perpendicular to its orientation. When the two particles carry the same charge, the dimer can rotate. An experimental realisation of a charged rigid dimer could be realised in macroscopic granules, see e.g.~\cite{scholz2021surfactants} for granular dimers. By highly charging a capacitor made of granules, with its plates separated by an isolating bar, one can create a rigid dimer carrying charge of 1 Coulomb or even more. The friction coefficient $\gamma$ for granules is estimated to be about $0.1$ kg/s~\cite{scholz2018inertial}. These values together with $B = 10$ T yield $\kappa = 100$. This implies that $\kappa$ of the order 
of one and above is achievable in macroscopic model dimers exposed to a high magnetic field. For $\kappa = 100$, a rigid dimer behaves as a fracton: it is practically immobile along its orientation while it can translate perpendicular to its orientation. 

It is interesting to note that the dynamics of charged Rouse dimers are very similar to point vortices in superfluids \cite{doshi2021vortices} and liquid systems \cite{aref1999four,eckhardt1988integrable,tophoj2008chaotic}. While a dimer with opposite charges is similar to the point vortex anti-vortex pairing, which undergoes translation, a dimer with same charges is analogous to a pair of vortices with same vorticity, which undergoes rotation. Our findings are also consistent with dynamics of systems dominated by Magnus forces. It has been recently shown that for mixtures of particles with opposite Magnus force, particle pairs can combine to form translating dipoles \cite{reichhardt2020dynamics}.

We have not considered explicit electrostatic interactions in our study beyond an effective description. These are expected to produce anharmonic internal modes which would, however, not change the dynamics qualitatively. Finally, the Rouse dimer considered in this paper should be extended in future work towards the Zimm model \cite{zimm1956dynamics} which incorporates hydrodynamic interactions between the monomers. For charged monomers, local charge neutrality of oppositely flowing counterions will strongly reduce these hydrodynamic
interactions in an external electric field \cite{long2001note, rex2008influence, nedelcu2013molecular, shendruk2012electrophoresis}, an effect which we would also expect for external magnetic fields. Recently, fluids characterized by an odd-viscosity tensor, for instance, collection of active chiral particles and actomyosin gels, have received considerable attention \cite{banerjee2017odd, markovich2021odd}. A collection of charged Rouse dimers under magnetic field can be regarded as an example of a fluid with odd-viscosity. It would be interesting to study the effect of odd-viscosity of such a fluid on the dynamics of a tracer particle. Unlike odd-elastic systems, where work can be extracted in strain-controlled quasistatic cycles \cite{hargus2021odd}, Lorentz force, by virtue of being perpendicular to velocity, cannot affect the energetics of a system. However, it will affect the relaxation dynamics of Rouse chains in a solvent and therefore the rheological properties, such as the loss modulus. We will explore this in a future study.

Rather than addressing a longstanding unsolved question in plasma and polyelectrolyte science, our research results explore an hitherto unknown area of plasmas and polyelectrolytes for high magnetic fields. Getting into this new regime requires high charges and high magnetic fields. We have shown that this is an unfamiliar area but is, in principle, within reach in complex plasmas.
Likewise, model polyelectrolytes obtained by linking highly charged colloids or granular particles together get into the regime where magneto-induced effects become visible when exposed to a large magnetic 
field. 
Finally, our results shed new light on the tunability of the dynamics via an external magnetic field due to the interplay of rotational and translational degrees of freedom. This may be an important stepping stone for constructing miniaturized engines \cite{martinez2017colloidal} and soft robots \cite{sitti2018miniature} for the future.

\section{Acknowledgments}
We thank Erik Kalz and Iman Abdoli for insightful discussions.

\section{Funding}
This work is supported by funds from the Deutsche Forschungsgemeinschaft (DFG) within the project SH 1275/3-1.

%\bibliography{apsbib}% Produces the bibliography via BibTeX.
%\printbibliography

\bibliography{apssamp}% Produces the bibliography via BibTeX.

\end{document}